# LUNAR IMPACT FLASHES FROM GEMINIDS, ANALYSIS OF LUMINOUS EFFICIENCIES AND THE FLUX OF LARGE METEOROIDS ON EARTH

**J. L. Ortiz[1], J. M. Madiedo[2, 3], N. Morales[1], P. Santos-Sanz[1] and F. J. Aceituno[1]**

[1] Instituto de Astrofísica de Andalucía, CSIC, Apt 3004, 18080 Granada, Spain.
[2] Facultad de Física, Universidad de Sevilla, Departamento de Física Atómica, Molecular y Nuclear, 41012 Sevilla, Spain.
[3] Facultad de Ciencias Experimentales, Universidad de Huelva. 21071 Huelva (Spain).

**ABSTRACT**

We analyze lunar impact flashes recorded by our team during runs in December 2007, 2011, 2013 and 2014. In total, 12 impact flashes with magnitudes ranging between 7.1 and 9.3 in V band were identified. From these, 9 events could be linked to the Geminid stream. Using these observations the ratio of luminous energy emitted in the flashes with respect to the kinetic energy of the impactors for meteoroids of the Geminid stream is estimated. By making use of the known Geminids meteoroid flux on Earth we found this ratio to be $2.1 \cdot 10^{-3}$ on average. We compare this luminous efficiency with other estimations derived in the past for other meteoroid streams and also compare it with other estimations that we present here for the first time by making use of crater diameter measurements. We think that the luminous efficiency has to be revised downward, not upward, at least for sporadic impacts. This implies an increase in the influx of kilogram-sized and larger bodies on Earth that has been derived thus far through the lunar impact flash monitoring technique.

## 1 INTRODUCTION

In previous works, light flashes produced by meteoroids impacting the night side of the Moon have been identified mostly during the peak activity of several major meteor showers (see e.g. Dunham et al. 2000, Ortiz et al. 2000, Yanagisawa and Kisaichi 2002, Cudnik et al. 2002, Ortiz et al. 2002, Yanagisawa et al. 2006, Cooke et al. 2006, Yanagisawa et al. 2008, Madiedo et al. 2015b). Luminous efficiencies of these impact processes have been published for Leonid and Perseid showers by comparing the impact rates on the Moon with predicted rates from works on meteor fluxes on Earth for those streams. This was done by making use of the formalism described in Bellot-Rubio et al (2000a). Unfortunately, the derived luminous efficiencies have uncertainties of around an order of magnitude. The luminous efficiency, which is the fraction of the kinetic energy of the impactor that is converted into visible light during the impact, is an important parameter. It plays a fundamental role when the lunar impact flash monitoring technique is employed to estimate the sporadic flux of interplanetary matter impacting the Earth-Moon system and the impact hazard for Earth (Ortiz et al. 2006, Madiedo et al. 2014a). In this paper we present luminous efficiencies determined for Geminid meteoroids. The impact speed of the Geminids on the Moon is lower than that of the Leonids and that of the Perseids, and on the other hand, the meteoroids linked to the Geminids are thought to be of asteroidal origin rather than of cometary origin, so their composition can be somewhat different to those of the Leonids and Perseids meteoroids. Therefore, obtaining the luminous efficiency from the Geminids is valuable.

The lunar impact flashes in the above mentioned papers and some impact flashes in Suggs et al. (2014) were linked to the corresponding meteoroid streams on the basis of the coincidence of the detection date with their peak activity, provided that the impact geometry was compatible (i.e., if meteoroids from the stream could impact at the location on the lunar surface where the flash was identified). But associating impact flashes to a given meteoroid source by using this simple approach does not provide any measure about the quality of this association, especially if the stream is a weak source of meteors on Earth. Establishing this association with a good enough confidence level is of importance in order to estimate different parameters, such as for instance the mass of the impactor, the size of the resulting crater, and the velocity-dependent (and so stream dependent) luminous efficiency. The problem of quantifying the confidence level of the association of Moon impact flashes and meteoroid streams has recently been addressed by Madiedo et al. (2015a,b). This was done by defining a parameter that estimates the probability that an impact flash is produced by a meteoroid from a given meteoroid stream or from the sporadic background. In section 4.1 we apply this parameter to our December campaigns and can conclude that the impact flashes of our December 2011 and 2014 runs were not caused by Geminids. In Section 4.2 we estimate the masses of the impactors. In Section 4.3 we derive luminous efficiencies for the 2007 and 2013 Geminids and compare them with other meteor showers. In section 4.4 we present another completely different means of estimating the luminous efficiency of lunar impacts by making use of the size of the impact crater associated to the largest impact flash detected so far (Madiedo et al. 2014a). This crater was recently identified in Lunar Reconnaissance Orbiter (LRO) images. We discuss the implications of our findings, especially concerning the influx of kilogram-sized (and larger) impactors on Earth. Finally, in section 4.5 estimates of crater diameters produced by the observed Geminids impacts are made.

## 2 INSTRUMENTATION AND METHODS

Our 2007 lunar impact flashes monitoring campaign was conducted at La Sagra Astronomical Observatory (latitude: 37.98283 ºN, longitude: 2.56571 ºW, height: 1520 m above the sea level), where two identical 0.36 m Schmidt-Cassegrain telescopes manufactured by Celestron were employed to monitor the same part of the night side of the Moon. Two telescopes were used in order to have duplicity of impact flash detections to distinguish true impact flashes from noise or cosmic ray hits in the detectors. This is the usual procedure that we follow to detect impact flashes unambiguously. As detectors we employed Mintron MTV12V1-EX CCD video cameras. The December 2011, 2013 and 2014 monitoring campaigns were conducted from the MIDAS survey observatory in Sevilla (latitude: 37.34611 ºN, longitude: 5.98055 ºW, height: 23 m above the sea level) where we operated two identical 0.36 m Schmidt-Cassegrain telescopes and also a smaller Schmidt-Cassegrain telescope with a diameter of 0.28 m. All of them are also manufactured by Celestron. These telescopes employed high-sensitivity CCD video cameras (model Watec 902H Ultimate, manufactured by Watec Corporation).

During the monitoring campaigns considered in this work the telescopes were aimed at a common area of the night side of the lunar surface, as already stated. The lunar terminator was kept outside the video images to prevent an excess of light from the illuminated side of the Moon. These telescopes were tracked at lunar rate, but they were manually recentered when necessary since perfect tracking of the Moon at the required precision is not feasible with this equipment in a fully automated way. Besides, f/3.3 focal reducers for Schmidt Cassegrain telescopes manufactured by Meade Corporation were also in order to increase the area monitored by the cameras. The Mintron MTV12V1-EX and Watec 902H Ultimate devices produce interlaced analog imagery compliant with the CCIR video standard. Thus, monochrome images are obtained at a rate of 25 frames per second (fps). These images were digitized and stored on hard disk with a resolution of 720x576 pixels by means of a video acquisition card connected to a PC computer. GPS time inserters were used to stamp time information on every video frame with an accuracy of 0.01 s. The MIDAS software (Madiedo et al. 2010, 2015a, 2015b) was employed to identify and analyze flashes in the recorded images.

## 3 OBSERVATIONS AND RESULTS

For the 2007 December run the impact geometry of meteoroids from the Geminid stream is shown in Figure 1. This monitoring took place during the waxing crescent phase, from 18:17 UT to 20:45 UT with a Moon age of ~4.7 days and a lunar disk illumination of ~23 %. The effective observing time was of 2.4 hours.

In December 2011, 2013 and 2014 the Moon could not be monitored by our telescopes during the peak activity of the Geminids. At the moments of peak activity of the Geminids the illumination of the lunar disk was of about 83% in 2011 and 94% in 2013. These values are well above the upper threshold illumination of about 60% suitable for this technique (Ortiz et al. 2006, Madiedo et al. 2014a). On 2014 December 14 the illumination of the lunar disk was of about 48%, but no monitoring could be organized around that date because of bad weather conditions. The December 2011 campaign was conducted between December 28d19h05m UT and December 30d23h55m UT (effective observing time of 7.9 hours), with a Moon age ranging between 4.1 and 6.1 days (lunar disk illumination between 18 and 37 %). In 2013 the monitoring period extended from December 4d17h55m UT to December 7d22h04m UT, with a Moon age ranging from 2.0 to 5.5 days (lunar disk illumination between 4 and 30 %). The effective observing time was of 10.4 hours. And finally, in 2014 a campaign was conducted between December 25d17h40m UT and December 29d00h10m UT (effective observing time of 13.4 hours), when the lunar disk illumination ranged from 17 to 51 % (Moon age between 4.0 and 7.5 days). Since these three campaigns took place far from the peak of any major meteor shower, the telescopes were aimed at an arbitrary area of the lunar disk.

For each telescope, the MIDAS software obtained a database containing impact flash candidates identified in the recorded images. From these we eliminated false detections, such as those produced by cosmic rays, by discarding those events that were recorded by only one telescope (Madiedo et al. 2015a). In this way we confirmed four impact flashes recorded in the 2007 campaign, one event in 2011, five in 2013 and two in 2014. These events are listed in Table 1 together with their V-band magnitude and the selenographic coordinates of the corresponding impact. As shown in the table, the duration of these flashes ranged between 0.02 and 0.12 s, and their magnitudes ranged between 7.1 and 9.3 in the V band. For the magnitude calibration we used the known V-band magnitudes of stars that appeared in the field of view of the telescope near the limb at some point during the observations. Typical errors in these calibrations are of the order of 0.3 mag. We also use the known Earthshine brightness as a function of lunar phase to double check the magnitude estimates obtained by the other method. This is the method employed in Ortiz et al. (2006). We typically obtained differences of around 0.3 mag, which we take as the uncertainty in our V magnitude calibrations.

# 4 DISCUSSION

## 4.1. Meteoroid source of the impact flashes

The problem of associating lunar impact flashes to a given meteoroid stream was recently addressed in Madiedo et al. (2015a,b). This association was performed by evaluating the probability parameter p, which measures the probability that an impact flash can be linked to a specific meteoroid source. That probability is given in equation (17) of Madiedo et al. (2015a). The probability depends on several parameters. $HR^{SPO}_{Earth}$ is the average hourly rate of sporadic meteors on Earth. We have considered here $HR^{SPO}_{Earth}=10$ meteors $h^{-1}$ (Dubietis and Arlt 2010). φ is the impact angle with respect to the local vertical and ν is a parameter defined for stream and sporadic meteoroids ($\nu^{ST}$ and $\nu^{SPO}$, respectively) which includes in the computations only those meteoroids with a kinetic energy above the threshold kinetic energy $E_m$ necessary to produce impact flashes detectable from Earth (Bellot-Rubio et al. 2000a; Madiedo et al. 2015a,b), so $\nu = K^{s-1}E_m^{1-s}$ where *K* is the kinetic energy ($m_o V^2/2$) of the threshold mass of meteoroids that produce meteors with magnitude 6.5 in the Earth's atmosphere, V is the impact velocity, and s is the mass index, which is related to the population index r (the ratio of the number of meteors with magnitude m+1 or less to the number of meteors with magnitude m or less) through the formula s=1+2.5 log(r). For sporadic meteoroids we have considered r=3.0 (Dubietis and Arlt 2010). The value of $m_o$, which depends on the geocentric velocity, can be estimated for sporadic and stream meteoroids from Eqs. (1) and (2) in Hughes (1987). The factors $\gamma^{ST}$ and $\gamma^{SPO}$ in Eq. (17) of Madiedo et al. (2015a) account for the fact that the gravitational focusing effect for stream and sporadic meteoroids on Earth and Moon is different. For sporadic meteoroids we have $\gamma^{SPO}$ = 0.77 (Ortiz et al. 2006). For stream meteoroids $\gamma^{ST}$ is estimated from Eq. (9) in Madiedo et al. (2015a). The factor σ takes into account that, in general, the meteoroid stream will be at a different distance from the Earth than from the Moon and so the density of meteoroids from that stream on both bodies would be different. Another important parameter is the ZHR (zenithal hourly rate) of shower meteors on Earth ($ZHR^{ST}_{Earth}$) at solar longitude λ (which corresponds to the time of detection of the impact flash) and is obtained as $ZHR^{ST}_{Earth} = ZHR^{ST}_{Earth}(max) \cdot 10^{-b|\lambda-\lambda_{max}|}$ (Jenniskens 1994), where $ZHR^{ST}_{Earth}(max)$ is the peak ZHR on Earth (corresponding to the date given by the solar longitude $\lambda_{max}$). The values for the peak ZHR for different meteoroid streams and the corresponding solar longitudes for these maxima can be obtained, for instance, from (Jenniskens 2006). However, for a more precise analysis it is preferable to measure this ZHR for the time corresponding to the detection of the impact flash. The above-defined threshold kinetic energy $E_m$ of the impactor corresponds to the minimum radiated energy $E_{r_m}$ on the Moon detectable from observations on Earth with our instruments. Both magnitudes are related by means of the luminous efficiency $E_{r_m}=\eta E_m$ . This minimum radiated energy, in turn, is related to the maximum visual magnitude for detectable impacts ($m_{max}$). With our experimental setup this magnitude is $m_{max} \approx 10$. $E_{r_m}$ and $E_m$, which depend on the Earth-Moon distance and so on the observing date, can be calculated by the procedure described in Madiedo et al. (2015a,b).

The value of ZHR that makes the probability p=0.5, versus $V_g$, is shown in Fig. 2 for three population indices. Equation (17) with Eqs. (15), (16), and (18) of Madiedo et al. (2015a) have been used to generate the figure. The plot has been obtained by setting $\sigma = 1$, $\varphi = 45$ º, and $\eta = 2 \cdot 10^{-3}$. This figure shows that, for a fixed value of the geocentric velocity, the minimum ZHR increases significantly with r. Thus, in order to claim an association between an impact flash and a meteoroid stream, the zenithal hourly rate must be higher for streams with high values of the population index. By following this approach, and for $\eta = 2 \cdot 10^{-3}$, the canonical value for the luminous efficiency employed by us in previous works (see, e.g., Ortiz et al. 2006, Madiedo et al. 2014a, Madiedo et al. 2015a,b), $ZHR_{min}$ yields ~ 6 meteors $h^{-1}$ for the Geminids (r = 2.5 and $V_g$ = 35 km $s^{-1}$), ~ 3 meteors $h^{-1}$ for the α-Capricornids ( r = 2.5 and $V_g$ = 25 km $s^{-1}$), ~ 2 meteors $h^{-1}$ for the Perseids (r = 2.0 and $V_g$ = 59 km $s^{-1}$), ~ 1 meteors $h^{-1}$ for the Quadrantids (r = 2.1 and $V_g$ = 41 km $s^{-1}$) and ~ 25 meteors $h^{-1}$ for the Leonids (r = 2.5 and $V_g$ = 71 km $s^{-1}$), for example. To obtain these threshold zenithal hourly rates we have assumed that the population index for meteoroids producing detectable impact flashes on the Moon is the same as the value of r corresponding to meteoroids from the same stream producing meteors in the atmosphere.

According to the International Meteor Oganization (IMO), the peak activity of the Geminid meteor shower in 2007 occurred at around 14:37 UT on December 14 with a maximum zenithal hourly rate of ~120 meteors $h^{-1}$ (http://www.imo.net/live/geminids2007/). Despite the fact that this peak took place under daylight conditions, the broad maximum gave rise to a ZHR of about 100 meteors $h^{-1}$ during our lunar impact flash monitoring campaign. This ZHR value was estimated from the recordings performed by our meteor observing stations in the South of Spain (Madiedo and Trigo-Rodríguez 2008, Madiedo 2014) and was also confirmed by IMO (http://www.imo.net/live/geminids2007/). Thus, since the impact geometry is favorable (Figure 1) and this ZHR is above the value of $ZHR_{min}$ ~ 6 meteors $h^{-1}$ estimated for Geminid meteoroids, it is clear that the impact flashes recorded in 2007 could be associated to the Geminids. To quantify the confidence level of this association, we have considered that the lunar impact flashes recorded during our 2007 monitoring campaign could be produced either by Geminid meteoroids or by meteoroids from the sporadic background. The minimum kinetic energy $E_m$ for detectability of impact flashes produced by Geminid meteoroids was calculated from Eq. (3) of Madiedo et al. (2015a) by assuming f = 2 and a luminous efficiency of $2\cdot10^{-3}$ and by taking into account that with our experimental setup the limiting visual magnitude for detectable impact flashes is of about 10. We have employed Eq. (17) of Madiedo et al. (2015a) to obtain $p^{GEM}$ by assuming $HR^{SPO}_{Earth}$ = 10 meteors $h^{-1}$ (Dubietis and Arlt 2010), r = 3.0 for sporadics (see e.g. Dubietis and Arlt 2010; Rendtel 2006), and $\sigma \sim 1$. The population index of the Geminids has been found to vary between 1.7 and 2.5 around the peak of this shower (Rendtel 2004, Arlt and Rendtel 2006). For these computations we have considered r = 2.5, which is the most unfavourable case according to the discussion above. For each event the impact angle φ was provided by the MIDAS software (Madiedo et al. 2015a). An average value of 17 km $s^{-1}$ has been assumed for the impact velocity on the Moon of sporadic meteoroids (Ortiz et al. 1999). The impact velocity for Geminid meteoroids was calculated by following the approach described in Madiedo et al. (2014a) and by taking into account that their geocentric velocity is 35 km $s^{-1}$ (Jenniskens 2006). However, the calculated values of the impact velocity differ from $V_g$ by at most 0.3 km $s^{-1}$, which is below the accuracy of the value taken for $V_g$. So, we have considered V = 35 km $s^{-1}$ for our computations. The calculated values for this probability parameter are listed in Table 2. These results show that the probability parameter, which ranges between 0.95 and 0.96, is well above 0.5 for the impact flashes identified in December 2007 and so these can be considered as produced by Geminid meteoroids.

When analyzing the most likely source of the meteoroids that produced the impact flashes recorded in 2011, 2013 and 2014, it must be mentioned that these events did not take place near the peak of the Geminids nor near any other major meteor shower. The event identified on 30 December 2011 took place within the activity period of the Quadrantids (December 28- January 12), but about 5 days away from the peak of this shower (January 4). Nevertheless, by the time of detection of this flash the ZHR of the Quadrantids was of about 1 meteor $h^{-1}$ according to the data obtained by our meteor observing stations, which fits the ZHR value of ~ 1 meteor $h^{-1}$ necessary to claim a link with this stream. Besides, the impact geometry for Quadrantid meteoroids was found to be unfavourable, since these particles could not impact the region on the Moon where the flash was identified. So, we conclude that the most likely scenario is that this event was produced by a sporadic meteoroid.

The impact flashes identified in December 2013 occurred within the activity period of the Geminids (which extends from December 4 to December 17), but between 7 and 9 days away from its peak (December 14). Nevertheless, the recordings performed by our meteor observing stations revealed that the Geminid shower was rich in bright meteors during the beginning of its activity period, which resulted in a population index of about 2.0 for meteors observed between December 4 and December 8 (see Figure 3). This population index is lower than the maximum value of r = 2.5 measured during the peak activity of the Geminid shower (Rendtel 2004, Arlt and Rendtel 2006). In 2013, however, an outburst of the Andromedid meteor shower with a peak activity of about 20 meteors $h^{-1}$ took place on December 7-8 (Green 2013). Before December 7-8, our meteor observing stations recorded an activity of around 1 meteor $h^{-1}$ or less for this shower. So, since the impact geometry of these flashes was found to be compatible with both meteoroid streams, to analyze the likely source of the impactors that produced the impact flashes recorded in 2013 we have considered that these could be produced by Geminid, Andromedid or sporadic meteoroids. For Andromedid meteors we have assumed a value of the population index of 3.0 (Wiegert et al. 2013). The corresponding probabilities calculated from Eq. (17) of Madiedo et al. (2015a) for the association with the Geminids and the Andromedids are listed in Table 3. These values show that the most likely scenario is that these flashes were produced by Geminid meteoroids, with probabilities ranging between 61 and 91 %. For the Andromedids, however, the probabilities are quite low, since these vary between 2 and 5 %.

Finally, we think the two flashes identified on 26 December 2014 were sporadic: both events took place by the end of the activity period of the Ursids, but with r = 3.0 and $V_g$ = 33 km $s^{-1}$, the minimum ZHR necessary to link a flash with the Ursid stream with a 50% probability yields ~ 30 meteors $h^{-1}$ (see Fig. 2), well above the usual activity level of this shower during its peak (around 10 meteors $h^{-1}$). But the activity of the Ursids experienced an outburst in 2014, reaching a maximum ZHR of about 50 meteors $h^{-1}$ around Dec. 23 (Brown and Jenniskens 2015), which could raise suspicion that the flashes could be due to this stream. Nevertheless, the activity around December 26 was of only about 2 meteors $h^{-1}$ according to the observations performed by our meteor observing stations. Thus we can safely conclude that the impact flashes on Dec 26, 2014 were sporadic.

## 4.2. Masses of the meteoroids that caused the observed impacts

The mass of the Geminid meteoroids that produced the impact flashes associated with this stream in Tables 2 and 3 can be obtained once that the luminous efficiency $\eta$ for these events has been estimated. This mass M is given by the equation

$$M=2E_r\eta^{-1}V^{-2} \quad \textbf{(1)}$$

where V is the impact velocity and $E_r$ is the radiated energy recorded by the telescopes. This energy is calculated by integrating the radiated power in time. To calculate the mass of sporadic meteoroids, we have followed the same procedure by employing $\eta=2\cdot10^{-3}$ (Ortiz et al. 2006). The results are summarized in Table 4, which also includes the diameter of these particles. To calculate this size a bulk density of 1.8 and 2.9 g $cm^{-3}$ has been assumed for sporadic and Geminid meteoroids, respectively (Babadzhanov and Kokhirova 2009). As can be seen, the mass of the impactors that produced the impact flashes analyzed in this work range between 5.3 and 223 g.

### 4.3. Luminous efficiency for the Geminid meteoroids

Moon impact flashes produced by Geminid meteoroids have been reported by other researchers (e.g. Yanagisawa et al. 2008), but luminous efficiencies were not derived from them. For the analysis of some parameters of the impactors, such as their mass, these authors assumed a value of the luminous efficiency of $2\cdot10^{-3}$.

The luminous efficiency for the impact flashes produced by Geminid meteoroids can be estimated by following the procedures described in Bellot Rubio et al. (2000a,b). The number N of expected impact flashes above an energy $E_d$ is given by:

$$N(E_d) = F(m_o)\Delta t\left(\frac{2f\pi R^2}{\eta m_o V^2}E_d\right)^{1-s} A \qquad \textbf{(2)}$$

where $\Delta t$ is the observing time, $E_d$ is the time-integrated optical energy flux of the flash observed on Earth, $F(m_o)$ is the flux of meteoroids on the Moon with mass higher than $m_o$, and A is the projected area of the observed lunar surface perpendicular to the Geminid meteoroid stream. In this analysis we have assumed f = 2. For the Geminids the impact speed is V = 35 km $s^{-1}$ and $m_o = 4.5 \cdot 10^{-7}$ kg. R, the distance of the Moon seen from Earth, is 384000 km on average. Besides, as a consequence of the different gravitational focusing effect for Moon and Earth defined by Eq. (9) in Madiedo et al. (2015a), the flux of meteoroids on Earth is higher than the flux of meteoroids on the Moon by a factor of 1.10.

From our 2007 monitoring we have obtained N = 4 impact flashes above an integrated energy flux $E_d = 7.0 \cdot 10^{-15}$ J m$^{-2}$. This is the value of $E_d$ for event #3 in Table 1, which is the lowest value of the integrated optical energy flux estimated for these 2007 Geminid flashes. Besides, $A = 1.3 \cdot 10^6$ km$^2$, R=384000 km at the time of the observations and $\Delta t = 2.4$ h. Since the 2007 monitoring took place at the peak activity of this shower, we have considered a value of 2.5 for the population index r. So, the mass index yields s = 2.0. According to the observations performed by our meteor observing stations, we have found that on Earth the average flux of faint Geminid meteoroids was $1.9 \cdot 10^{-2}$ meteors km$^{-2}$ h$^{-1}$. In this way the value of the luminous efficiency obtained by using equation (**2**) yields $\eta = 1.\mathbf{8} \cdot 10^{-3}$. See Figure 4 for an illustration on how equation (2) reproduces the available data.

We have repeated this procedure for the N = 5 Geminid impact flashes imaged in 2013 (Table 3). The minimum energy flux for these events is $E_d = 2.6 \cdot 10^{-14}$ J m$^{-2}$. In this case, as discussed above, we have considered r = 2.0. The total monitoring time was $\Delta t = 10.4$ h, with $A = 1.3 \cdot 10^6$ km$^2$ and R=361000 km at the time of the observations. The averaged flux of Geminid meteoroids on Earth during the impact flashes monitoring period, as determined from the data obtained by our meteor observing stations, was $1\cdot10^{-3}$ meteors $km^{-2}$ $h^{-1}$. With these values, a luminous efficiency of $2.4\cdot10^{-3}$ is derived from Eq. (2). See Figure 4 for an illustration on how equation (2) reproduces the available data.

By averaging the value of the luminous efficiency determined from the flashes identified in 2007 and 2013 we obtain $\eta=2.1\cdot10^{-3}$. This efficiency is very close to the canonical $\eta=2\cdot10^{-3}$ value used above to estimate the probability parameter. This $\eta=2\cdot10^{-3}$ value is the one obtained in the past for the Leonids (Bellot Rubio et al. 2000a, Ortiz et al. 2002) and is also close to the $1.8\cdot10^{-3}$ efficiency determined for the Perseids (Madiedo et al. 2015a) and not far from the $3.4\cdot10^{-3}$ efficiency of the alpha Capricornids (Madiedo et al. 2015a) despite the different impact speeds involved**.** From all these measurements and contrary to initial expectations, it appears that the luminous efficiency does not increase with impact speed or the dependence is weak at these high speeds.

## 4.4. Constraints on the luminous efficiency from the measured size of the new crater on the Moon formed from the September 11th, 2013 large impact blast.

On September 11th 2013 a very bright impact flash, the brightest impact flash ever detected, was observed from our lunar monitoring systems (Madiedo et al. 2014a). This impact gave rise to a large enough crater to be easily identifiable by the Lunar Reconnaissance Orbiter (LRO) cameras. Indeed, this was the case and a crater of 34 m in diameter was found **at** the impact site coordinates. This was released by the LRO team through the internet at http://lroc.sese.asu.edu/posts/810

This offers the possibility of comparing the actual crater size with theoretical computations of the crater size based on the calculated energy deposition and on plausible values of certain parameters. To estimate the diameter of the craters produced on the Moon from an impact we have employed the following equation (Gault 1974, Melosh 1989), which is suitable for small craters (diameter < 100m) and has been used by other researchers in the field to estimate impact crater sizes on the Moon (e.g. Suggs et al. 2014):

$$D = 0.25\rho_p^{1/6}\rho_t^{-0.5}E^{0.29}(\sin\theta)^{1/3} \qquad \textbf{(3)}$$

where D is the crater diameter, θ the impact angle with respect to the horizontal, E the kinetic energy of the meteoroid, and $\rho_p$ and $\rho_t$ the impactor and target bulk densities, respectively. In this relationship these quantities are expressed in mks units. For sporadic meteoroids we have taken $\rho_p$ =

1800 kg $m^{-3}$ (e.g. Babadzhanov and Kokhirova 2009) and $\theta$ = 45°. For the density of the target we take 1600 kg $m^{-3}$, which is slightly higher than that of the lunar regolith (expected to be between 1300 to 1500 kg $m^{-3}$) but smaller than that of the lunar megaregolith. We do not expect that the crater depth could be enough to have reached the megaregolith because the regolith thickness is typically 5 to 15 m depending on the terrain (see e.g. Han et al. 2014 for a review of densities and thicknesses of the lunar regolith and megaregolith), but we take here a density of 1600 kg $m^{-3}$ for the target to account for the possibility that the crater could have reached the megaregolith. Using the above mentioned values and the kinetic energy of the impactor derived in Madiedo et al. (2014a) we come up with a diameter of 27 m, which is not close enough to the actual value of 34 m measured by LRO. This may probably mean that the kinetic energy has been somewhat underestimated in Madiedo et al. (2014a). In that work a luminous efficiency $\eta=2\cdot10^{-3}$ was used to derive the kinetic energy. If we use a value of $\eta=7\cdot10^{-4}$ we come up with the right kinetic energy so that the crater diameter becomes exactly 34m, as observed. There is also the possibility that the impactor that caused the flash was denser than the value of 1800 kg $m^{-3}$ used here. If we use a density of 3000 kg $m^{-3}$ typical of a chondrite meteorites while keeping $\eta=2\cdot10^{-3}$, we come up with a crater size of 30m, which is closer to the correct value, but even in this case, the luminous efficiency has to be decreased at least to $1.1\cdot10^{-3}$ to produce the needed kinetic energy. So this crater diameter estimate calls for a lower luminous efficiency than that derived using meteoroid streams as calibrators.

In the original Gault (1974) paper it is not stated whether the crater diameter used in equation (3) is the rim to rim diameter or the apparent diameter. These two diameters can differ by about 30% according to Housen et al. (1983). In Melosh (1989), the Gault (1974) equation is written for D_ap, meaning apparent diameter. If we interpret that we should use the apparent diameter instead of the rim to rim diameter, the measured diameter is basically coincident with the calculation using just a 10% smaller luminous efficiency than the nominal luminous efficiency $\eta=2\cdot10^{-3}$. In order to shed more light on the issue we have used another expression of crater diameter based on other more recent research works. In the review paper by Holsapple (1993) the following expression is used for the rim to rim radius of vertical impacts, in meters:

$$R = 10.14 G^{-0.17} a^{0.83} V^{0.34} \qquad (4)$$

Where a is the impactor radius (in meters), G is the gravity acceleration at the target surface (in units of Earth's gravity acceleration g) and V is the impact speed. Note that the speed must be entered in km/s in this equation. If we account for a non vertical impact angle using the same dependence as shown in equation (3), we come up with a rim to rim impact diameter of 29m for the impact of 11 September 2013, using the same parameters as above and a typical impact speed of a sporadic impactor on the Moon (17 km/s). This diameter estimate is close to that obtained using equation (3) if it is rim to rim diameter, not apparent diameter. The computed rim to rim diameter of 29m is smaller than the observed one and again the easiest explanation is that a larger kinetic energy is needed in the equation, which calls for a downward revision of the luminous efficiency. The required value is $\eta=1.1\cdot10^{-3}$. Either that or the particular impactor was denser than a typical sporadic meteoroid or the impact angle was vertical.

To also put further constraints on the luminous efficiency, we can take advantage of the diameter of a smaller new impact crater reported in Robinson et al. (2015) that was formed on March 17, 2013, from an impact that caused a bright impact flash (Suggs et al. 2014). Using the impact kinetic energy of $5.4\cdot10^{9}$ J reported by Suggs et al. (2014) and the same values of the density parameters and impact angle as for the September 11th impact, we come up with a crater diameter of only 12m, whereas the measured diameter is >50% larger (18.8m according to Robinson et al. 2015). In order to get a diameter of 18.8 m we have to increase the kinetic energy by a factor of ~3.8. Because the Suggs et al. (2014) work used a luminous efficiency of $1.4\cdot10^{-3}$ a reduction of this luminous efficiency by a factor of ~3.8 would be needed, which would mean an $\eta$ value of $4\cdot10^{-4}$. This is far too small, much smaller than the already considerably small bound of $7\cdot10^{-4}$ determined in the previous paragraph from the 2013 September 11$^{th}$ bright flash. If we use the rim to rim crater diameter from equation **(3)** we obtain again a diameter that is too small compared to the observations. Note that in this case we used a speed of 17 km/s instead of the speed of the Virginids because the qualitative link of the 11 March 2013 lunar impact to the Virginids stated in Suggs et al. (2014) does not hold when we use our quantitative approach of calculating the probability parameter.

Hence, the kinetic energy estimated by Suggs et al. (2014) for the impact flash is too low. The too low kinetic energy in the work by Suggs et al. (2014) can be explained because their emitted energy estimate is too low. This can be explained because in their expression of the radiated energy they use a width of the passband that is too small for unfiltered observations. The width of the passband used by the authors is for an R filter, but their reported Lunar observations were not obtained through an R filter. When a correct band width is used, the emitted energy increases by around a factor of nearly 4, and so does the kinetic energy. Then, the diameter of the crater becomes exactly 18.8 m, as observed.

Another clear evidence of the too low emitted energy calculations in Suggs et al. (2014) is obtained by comparing the $8.9 \cdot 10^3$ J emitted energy for a magnitude 9.5 impact flash (impact# 100 in table 1 of Suggs et al. 2014), with the emitted energy of the magnitude 9.5 Perseid impact flash reported in Yanagisawa et al. (2006), which is $4.1 \cdot 10^4$ J according to these authors in page 493. This is a discrepancy of a factor of 4.6. A small part of the discrepancy may arise from the fact that Suggs et al. (2014) refer to R magnitude whereas the Yanagisawa et al. (2006) magnitude is in V band. If we use a V-R color difference of around 0.5 mag for typical impact flashes, which may be possible, depending on the unknown impact plume temperature, we can compare the Yanagisawa et al. (2006) emitted energy with that of an R magnitude 9 impact flash in table 1 of Suggs et al. (2014) such as impact #22. In this case the difference of emitted energy is a factor 3.1. In summary, the emitted energy computations in Suggs et al. (2014) are too low by a factor that can range from 4.6 to 3.1. The disagreement in emitted energy by Suggs et al. (2014) for impacts of identical magnitude is not only with Yanagisawa et al. (2006) but also with the emitted energies reported in Ortiz et al. (2002) for similar magnitudes, despite these two latter works used completely different calibrations schemes.

Besides, when a factor ~4 is applied to the emitted energies in Suggs et al. (2014), the cumulative number of objects colliding with the Earth per year as a function of their kinetic energy (in their figure 9) agrees reasonably well with the impact rate measurements in Ortiz et al. (2006) provided that the same luminous efficiencies are used in the two studies.

In summary, the crater sizes give a hint for luminous efficiencies in the range of $\sim 7\cdot 10^{-4}$ to $\sim 2\cdot 10^{-3}$. Although the luminous efficiencies derived from lunar crater diameters are considerably uncertain because of the limitations in the scaling laws, they are somewhat smaller than the canonical $2\cdot 10^{-3}$ value used often, which was uncertain by around an order of magnitude (because it critically depends on the population index of the Leonid stream used as calibrator).

This implies that the flux of kg-sized bodies on Earth obtained from the lunar monitoring technique is considerably increased with respect to the best-fit straight line in Brown et al. (2002) plots. As shown in figure 5 for the intermediate $\eta$ between $7 \cdot 10^{-4}$ and $2 \cdot 10^{-3}$ , the Ortiz et al. (2006) impact rate and that of Madiedo et al. (2014a), depart considerably from the Brown et al. (2002) flux. This is not very surprising, given that Brown et al. (2013) have already revised their impact-rate for Chelyabinsk-like impactors by at least an order of magnitude upward with respect to the Brown et al. (2002) values, and for somewhat smaller impactors than the Chelyabinsk projectile, this can also be the case. These impact rates are consistent with the rates reported by Ceplecha et al. (1996, 2001) based on large bolide fluxes measured in the past, and often neglected by the scientific community in recent years. Furthermore, this is also consistent with superbolide fluxes reported in Madiedo et al. (2014b) based on the detection of 3 very bright bolides in a relatively short time span.

Whether or not the luminous efficiency of lunar impacts depends on the size and mass of the impactor or on its kinetic energy remains to be investigated with more data. The results shown in this paper based on the Geminids apply to small impactors, whereas the results based on the two lunar craters apply to considerably larger bodies and larger energies. The integrated luminous efficiency of meteors on Earth seems to be energy dependent (e.g. Brown et al. 2002), with higher efficiencies for higher kinetic energies. This might also be the case for lunar impact flashes but the luminous efficiency based on impact crater sizes seems to imply the opposite trend, although another explanation for this is a lower luminous efficiency for impacts of lower speed. However, no dependence of luminous efficiency with impact speed is observed in meteoroid streams of speeds ranging from 71 to 35 km/s, as shown in this paper. Possibly, one of the best strategies to study this and other possibilities would be to find new impact craters of different sizes by means of LRO imagery of lunar sites with reported impact flashes of different magnitudes and for different meteoroid streams so that a complete and statistically significant sample can be built.

## 4.5. Estimates of crater sizes of the observed impacts

Using expressions (3) and (4) for the Geminds reported here, the resulting crater diameters are listed in Table 4. For the calculations we have used $\rho_t = 1.6$ g cm$^{-3}$ as already mentioned. For Geminid and sporadic meteoroids we have taken again $\rho_p = 2.9$ g cm$^{-3}$ and $\rho_p = 1.8$ g cm$^{-3}$, respectively (Babadzhanov and Kokhirova 2009). These small craters, with diameters ranging between 0.55 and 1.44 m, would be hard to identify and measure by LRO.

## 5 CONCLUSIONS

In this work we report impact flashes observed in December runs for different years. The majority of these impacts can be attributed to Geminid meteoroids using a probability parameter for the association of impact flashes with active meteoroid streams. From the brightness of these impact flashes the luminous efficiency has been derived using the observed meteor fluxes and population indices on Earth as the main parameters needed for the impact energy calibration. The average luminous efficiency is $2.1 \cdot 10^{-3}$. This is consistent with other luminous efficiencies of other meteoroid streams published in the past and indicates that the luminous efficiency **does** not strongly depend on impact speed at least for speeds higher than 35 km/s**.** To further constraint the luminous efficiency, we have used for the first time crater size determinations obtained from LRO images of the Moon at the locations of two very bright impact flashes. Using the measured diameters and the theoretical diameters, which depend on the kinetic energy of the impactor, constraints on the luminous efficiency can be obtained. These constraints suggest a downward revision of the luminous efficiency, at least for sporadic impactors. This has an important effect on the rate of impacts on Earth of kilogram-sized impactors (and somewhat larger) that is derived using the lunar flash monitoring technique. The influx of bodies on Earth from the lunar impact monitoring technique is higher than that reported in Brown et al. (2002) but consistent with fluxes reported by Ceplecha (1996, 2001) based on the observations of large bolides in the past, often neglected in the current scientific literature, and also consistent with the somewhat crude bolide flux estimations based on three superbolides reported in Madiedo et al. (2014b). This is also consistent with the upward revised flux of impacts of Chelyabinsk size (Brown et al. 2013) if the increase they report is not only for decameter-sized impactors, but also for smaller bodies.

**ACKNOWLEDGEMENTS**

Funds from the Proyecto de Excelencia de la Junta de Andalucía, J.A. 2012-FQM1776 and from FEDER are acknowledged.

**REFERENCES**

Arlt R., Rendtel J., 2006, MNRAS, 367, 1721.

Babadzhanov P.B. and Kokhirova G.I., 2009, A&A, 495, 353.

Bellot Rubio L.R., Ortiz J.L., Sada P.V., 2000a, ApJ, 542, L65.

Bellot Rubio L.R., Ortiz J.L., Sada P.V., 2000b, Earth Moon Planets 82–83, 575.

Brown, P., Spalding, R. E., ReVelle, D. O., Tagliaferri, E., Worden, S. P. 2002. Nature, 420, 294-296.

Brown, P. G. et al. 2013. Nature, 503, 238-241.

Brown P., Jenniskens P., 2015, Central Bureau Electronic Telegrams, 4041, 1.

Ceplecha, Z. 1996, Astronomy and Astrophysics, v.311, p.329-332

Ceplecha, Z., 2001, in Marov M. Y., Rickman H., eds, Astrophysics and Space Science Library, Vol. 261, Collisional processes in the solar system. Kluwer, Dordrecht, p. 35

Cooke W.J., Suggs R.M., Swift W.R., 2006, Lunar Planet. Sci. 37. Abstract 1731.

Cudnik B.M., Dunham D.W., Palmer D.M., Cook A.C., Venable J.R., Gural P.S., 2002, Lunar Planet. Sci. 33. Abstract 1329C.

Dubietis A, Arlt R, 2010, EMP, 106, 105.

Dunham, D. W., Cudnik, B., Palmer, D. M., Sada, P. V., Melosh, J., Frankenberger, M. Beech R., Pellerin, L., Venable, R., Asher, D., Sterner, R., Gotwols, B., Wun, B., Stockbauer, D., 2000, Lunar Planet. Sci. 31. Abstract 1547

Gault D.E., 1974, In: R. Greeley, P.H. Schultz (eds.), A primer in lunar geology, NASA Ames, Moffet Field, p. 137.

Green D.W.E., 2013, Central Bureau Electronic Telegrams, 3741, 1.

Han, S.-C., Schmerr, N., Neumann, G., Holmes, S., 2014. Geophysical Research Letters, 41, 1882.

Housen, K. R., Schmidt, R. M ., Holsapple, K. A., 1983. J. Geophys. Res. 88, 2485-2499

Holsapple, K. A., 1993. Annu. Rev. Earth Planet. Sci. 21, 333

Jenniskens P., 1994, A&A, 287, 990.

Jenniskens P., 2006, Meteor Showers and their Parent Comets. Cambridge University Press.

Madiedo J. M., 2014, Earth, Planets and Space, 66, 70.

Madiedo J. M., Trigo-Rodríguez J. M., 2008, Earth Moon Planets, 102, 133.

Madiedo J. M., Trigo-Rodríguez J. M., Ortiz, J. L., Morales, 2010, Advances in Astronomy, 2010, article id. 167494. doi:10.1155/2010/167494

Madiedo J.M., Ortiz J.L., Morales N., Cabrera-Caño J., 2014a, MNRAS, 439, 2364.

Madiedo J.M. et al., 2014, Icarus, 233, 27.

Madiedo J.M., Ortiz J.L., Morales N., Cabrera-Caño J., 2015a, PSS, 111, p. 105-115.

Madiedo J.M., Ortiz J.L., Organero F., Ana-Hernández L., Fonseca F., Morales N., Cabrera-Caño J., 2015b, A&A, 577, id. A118, 9 pp.

Melosh H.J., 1989. Impact Cratering: A Geologic Process. Oxford Univ. Press, New York.

Ortiz J.L., Aceituno F.J., Aceituno J., 1999. A&A, 343, L57.

Ortiz J.L., Sada P.V., Bellot Rubio L.R., Aceituno F.V., Aceituno J., Gutierrez P.J., Thiele U., 2000, Nature, 405, 921.

Ortiz J.L., Quesada J.A., Aceituno J., Aceituno F.J., Bellot Rubio L.R., 2002, ApJ, 576, 567.

Ortiz J.L. et al., 2006, Icarus, 184, 319.

Rendtel J., 2004, Earth Moon Planets, 95, 27.

Robinson M.S., Boyd A.K., Denevi B.W.; Lawrence S.J., McEwen A.S., Moser D.E., Povilaitis R.Z., Stelling R.W., Suggs R.M., Thompson S.D., Wagner R.V., 2015, Icarus, 252, 229.

Suggs, R. M., Moser, D. E., Cooke, W. J., & Suggs, R. J., 2014, Icarus, 238, 23

Wiegert P., Brown P.G., Weryk R.J., Wong D.K., 2013, AJ, 145, 70.

Yanagisawa M., Kisaichi N., 2002, Icarus, 159, 31.

Yanagisawa M., Ohnishi K., Takamura Y., Masuda H., Ida M., Ishida M., 2006, Icarus, 182, 489.

Yanagisawa M., Ikegami H., Ishida M., Karasaki H., Takahashi J., Kinoshita K., Ohnishi K, 2008, Meteoritics and Planetary Science Supplement, Vol. 43, paper id. 5169.

# TABLES

| Flash # | Date and time (UTC) | Selenographic coordinates | mag | τ (s) | $E_d$ (J m$^{-2}$) |
|---|---|---|---|---|---|
| 1 | 14/Dec/2007 19:18:06 | Lat: 7.4±0.2 ºS Lon: 51.2±0.2 ºW | 9.2±0.3 | 0.02 | $7.7 \cdot 10^{-15}$ |
| 2 | 14/Dec/2007 19:28:48 | Lat: 17.6±0.3 ºN Lon: 58.2±0.3 ºW | 8.2±0.3 | 0.10 | $9.6 \cdot 10^{-14}$ |
| 3 | 14/Dec/2007 19:50:57 | Lat: 5.5±0.2 ºS Lon: 4.4±0.9 ºW | 9.3±0.3 | 0.02 | $7.0 \cdot 10^{-15}$ |
| 4 | 14/Dec/2007 20:42:57 | Lat: 25.3±0.5 ºN Lon: 38.2±0.5 ºW | 7.2±0.3 | 0.04 | $9.7 \cdot 10^{-14}$ |
| 5 | 30/Dec/2011 21:00:30 | Lat: 12.9±0.2 ºN Lon: 27.6±0.2 ºW | 8.5±0.3 | 0.04 | $2.9 \cdot 10^{-14}$ |
| 6 | 5/Dec/2013 18:29:41 | Lat: 2.3±0.2 ºS Lon: 11.4±0.2 ºW | 8.1±0.3 | 0.06 | $6.3 \cdot 10^{-14}$ |
| 7 | 5/Dec/2013 19:00:06 | Lat: 9.2±0.2 ºS Lon: 50.2±0.2 ºW | 8.8±0.3 | 0.06 | $3.3 \cdot 10^{-14}$ |
| 8 | 5/Dec/2013 19:03:14 | Lat: 12.0±0.2 ºS Lon: 38.3±0.2 ºW | 7.5±0.3 | 0.10 | $1.8 \cdot 10^{-13}$ |
| 9 | 6/Dec/2013 18:56:13 | Lat: 24.2±0.2 ºS Lon: 31.6±0.2 ºW | 8.6±0.3 | 0.04 | $2.6 \cdot 10^{-14}$ |
| 10 | 7/Dec/2013 19:31:06 | Lat: 14.6±0.2 ºS Lon: 10.6±0.2 ºW | 7.1±0.3 | 0.12 | $3.2 \cdot 10^{-13}$ |
| 11 | 26/Dec/2014 18:42:15 | Lat: 20.5±0.2 ºN Lon: 75.1±0.4 ºW | 8.1±0.3 | 0.04 | $4.2 \cdot 10^{-14}$ |
| 12 | 26/Dec/2014 20:52:02 | Lat: 2.4±0.2 ºS Lon: 63.4±0.2 ºW | 7.3±0.3 | 0.12 | $2.6 \cdot 10^{-13}$ |

Table 1. Characteristics of the confirmed lunar impact flashes discussed in the text. τ: flash duration; mag: peak magnitude of the flash in V band; $E_d$: time-integrated optical energy flux of the flash observed on Earth.

| Date and time (UTC) | $\varphi$ (°) | $ZHR^{ST}_{Earth}$ ($h^{-1}$) | r | $m_o$x$10^{-7}$ (kg) | $V_g$ (km $s^{-1}$) | V (km $s^{-1}$) | $E_m$x$10^6$ (J) | $\nu^{SPO}$ x$10^{-5}$ | $\nu^{GEM}$ x$10^{-5}$ | $p^{GEM}$ | Stream |
|---|---|---|---|---|---|---|---|---|---|---|---|
| 14/Dec/2007 19:18:06 | 18 | 100 | 2.5 | 4.5 | 35 | 35 | 3.34 | 4.1 | 8.7 | 0.96 | GEM |
| 14/Dec/2007 19:28:48 | 26 | 100 | 2.5 | 4.5 | 35 | 35 | 3.34 | 4.1 | 8.7 | 0.95 | GEM |
| 14/Dec/2007 19:50:57 | 28 | 100 | 2.5 | 4.5 | 35 | 35 | 3.34 | 4.1 | 8.7 | 0.95 | GEM |
| 14/Dec/2007 20:42:57 | 15 | 100 | 2.5 | 4.5 | **35** | 35 | 3.34 | 4.1 | 8.7 | 0.96 | GEM |

Table 2. Values of the parameters employed to test the association of impact flashes recorded in 2007 with the Geminids.

| Date and time (UTC) | $\varphi$ (°) | $ZHR^{ST}_{Earth}$ ($h^{-1}$) | r | $m_o$x$10^{-7}$ (kg) | $V_g$ (km $s^{-1}$) | V (km $s^{-1}$) | $E_m$x$10^6$ (J) | $\nu^{SPO}$ x$10^{-5}$ | $\nu^{ST}$ x$10^{-4}$ | $p^{ST}$ | Stream |
|---|---|---|---|---|---|---|---|---|---|---|---|
| 5/Dec/2013 | 15 | 1 | 2.0 | 4.5 | 35 | 35 | 3.13 | 4.5 | 8.9 | 0.68 | GEM |
| 18:29:41 | 75 | 1 | 3.0 | 127 | 16 | 16 | | | 1.2 | 0.02 | AND |
| 5/Dec/2013 | 24 | 1 | 2.0 | 4.5 | 35 | 35 | 3.13 | 4.5 | 8.9 | 0.65 | GEM |
| 19:00:06 | 48 | 1 | 3.0 | 127 | 16 | 16 | | | 1.2 | 0.05 | AND |
| 5/Dec/2013 | 13 | 1 | 2.0 | 4.5 | 35 | 35 | 3.13 | 4.5 | 8.9 | 0.66 | GEM |
| 19:03:14 | 51 | 1 | 3.0 | 127 | 16 | 16 | | | 1.2 | 0.04 | AND |
| 6/Dec/2013 | 15 | 2 | 2.0 | 4.5 | 35 | 35 | 3.13 | 4.5 | 8.9 | 0.79 | GEM |
| 18:56:13 | 45 | 1 | 3.0 | 127 | 16 | 16 | | | 1.2 | 0.03 | AND |
| 7/Dec/2013 | 27 | 10 | 2.0 | 4.5 | 35 | 35 | 3.13 | 4.5 | 8.9 | 0.91 | GEM |
| 19:31:06 | 76 | 20 | 3.0 | 127 | 16 | 16 | | | 1.2 | 0.05 | AND |

Table 3. Values of the parameters employed to test the association of impact flashes recorded in 2013 with the Geminids and the Andromedids.

| Flash # | Date and time (UTC) | Meteoroid stream | M (g) | $D_p$ (cm) | D (m) |
|---|---|---|---|---|---|
| 1 | 14/Dec/2007 19:18:06 | GEM | 5.8±0.5 | 1.5±0.1 | 0.58±0.02 |
| 2 | 14/Dec/2007 19:28:48 | GEM | 73±5 | 3.6±0.1 | 1.19±0.03 |
| 3 | 14/Dec/2007 19:50:57 | GEM | 5.3±0.5 | 1.5±0.1 | 0.55±0.02 |
| 4 | 14/Dec/2007 20:42:57 | GEM | 74±6 | 3.6±0.1 | 1.23±0.04 |
| 5 | 30/Dec/2011 21:00:30 | SPO | 25±3 | 3.0±0.1 | 0.75±0.03 |
| 6 | 5/Dec/2013 18:29:41 | GEM | 44±4 | 3.0±0.1 | 1.05±0.03 |
| 7 | 5/Dec/2013 19:00:06 | GEM | 23±2 | 2.4±0.1 | 0.86±0.02 |
| 8 | 5/Dec/2013 19:03:14 | GEM | 128±10 | 4.4±0.1 | 1.44±0.03 |
| 9 | 6/Dec/2013 18:56:13 | GEM | 18±2 | 2.2±0.1 | 0.81±0.02 |
| 10 | 7/Dec/2013 19:31:06 | GEM | 223±20 | 5.2±0.2 | 1.64±0.03 |
| 11 | 26/Dec/2014 18:42:15 | SPO | 29±3 | 3.1±0.1 | 0.78±0.03 |
| 12 | 26/Dec/2014 20:52:02 | SPO | 180±16 | 5.7±0.2 | 1.33±0.03 |

Table 4. Meteoroid source, impactor mass M, impactor diameter $D_p$, and crater size D, derived for the lunar impact flashes analyzed in the text.

# FIGURES

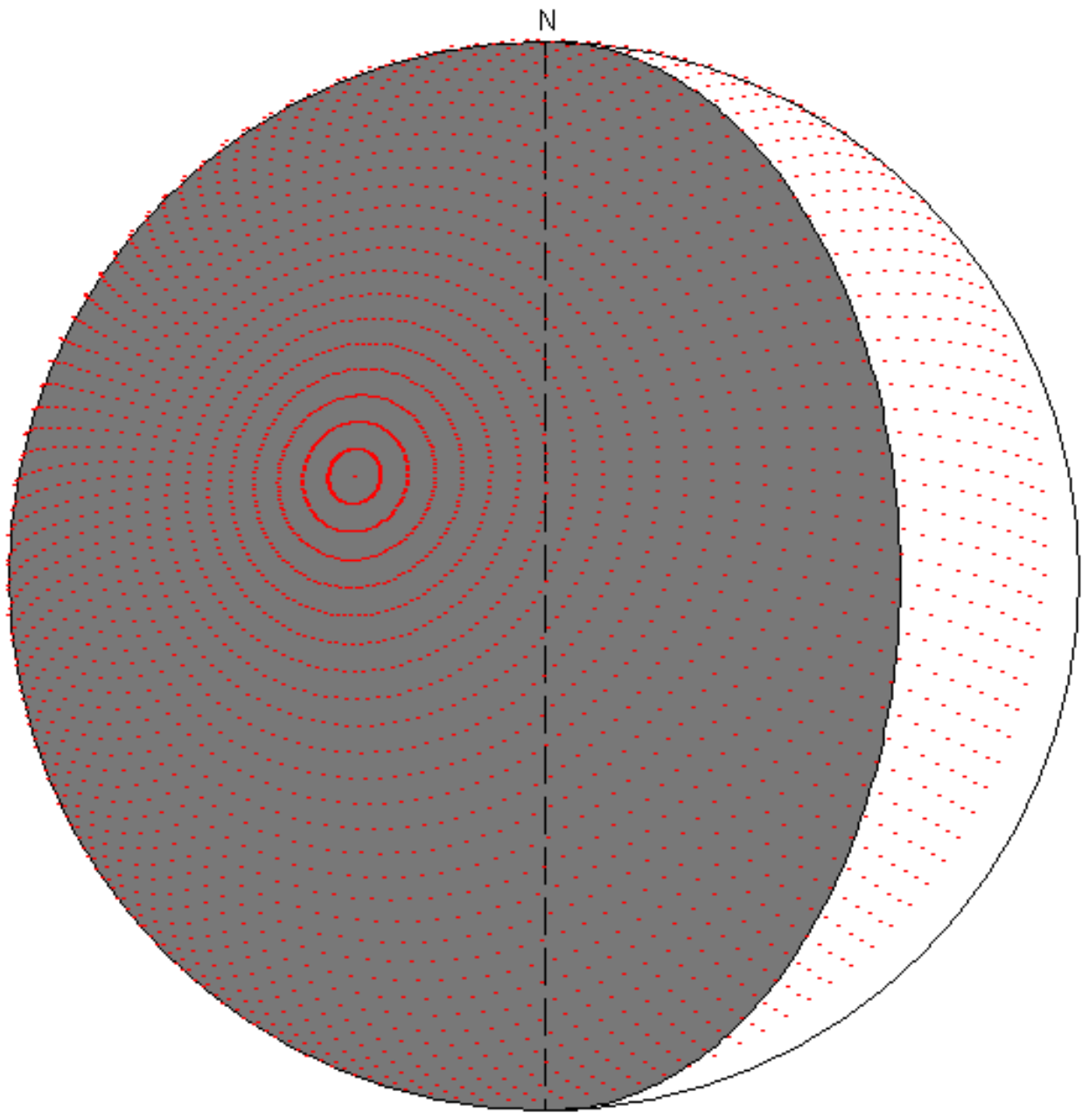

Figure 1. The lunar disk on 2007 December 14. White region: area illuminated by the Sun. Gray region: night side. Dotted region: area where Geminid meteoroids could impact.

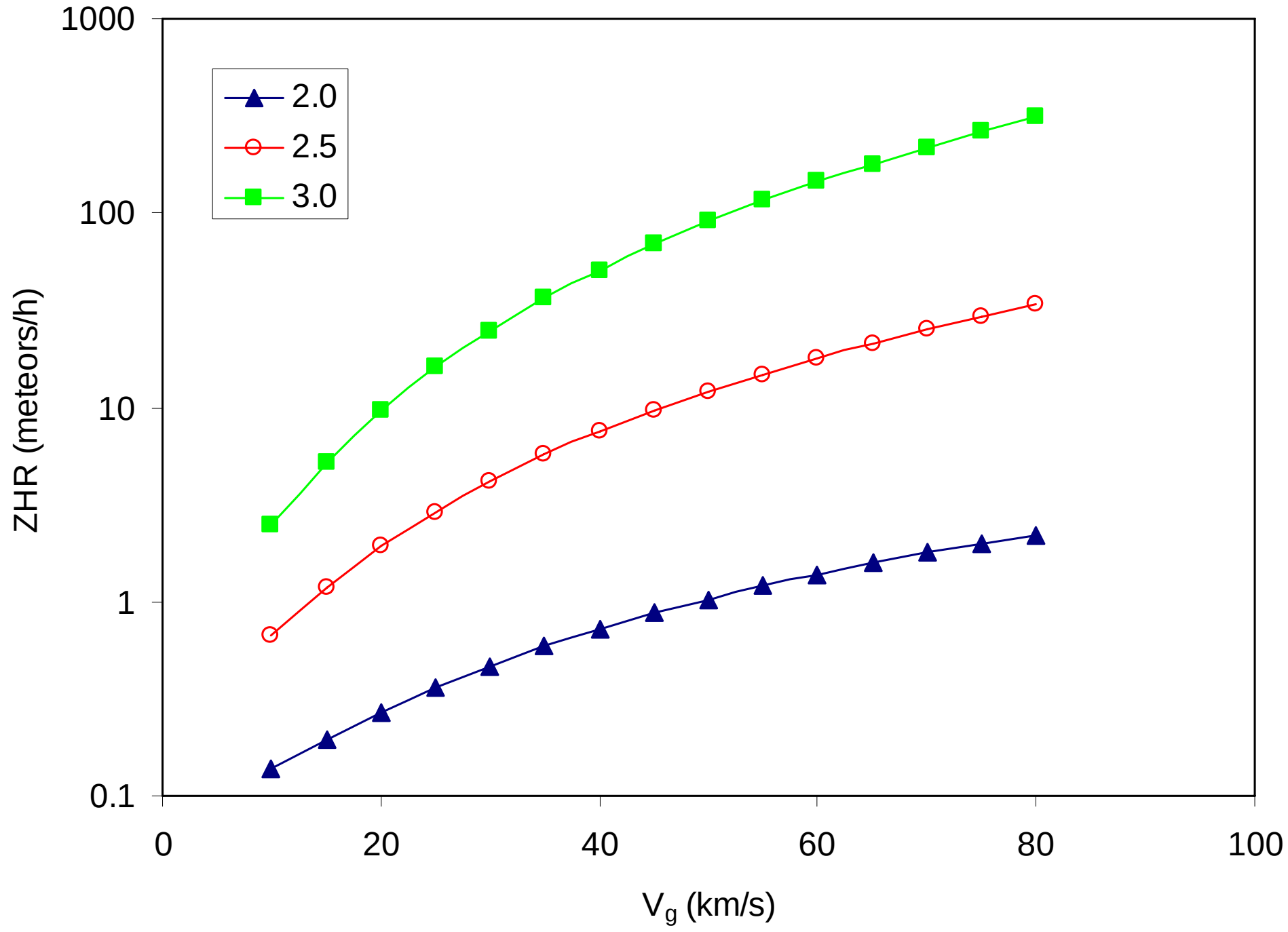

Figure **2**. Zenithal hourly rate (ZHR) of a stream that, according to Eq. (17) of Madiedo et al. (2015a), allows establishing a link between an impact flash and a meteoroid stream with a 50% probability, verus the geocentric velocity $V_g$ of the impactors. The calculations have been performed by considering population indices of 2.0, 2.5 and 3.0 (in blue, red and green color, respectively), and by setting $\eta = 2\cdot10^{-3}$.

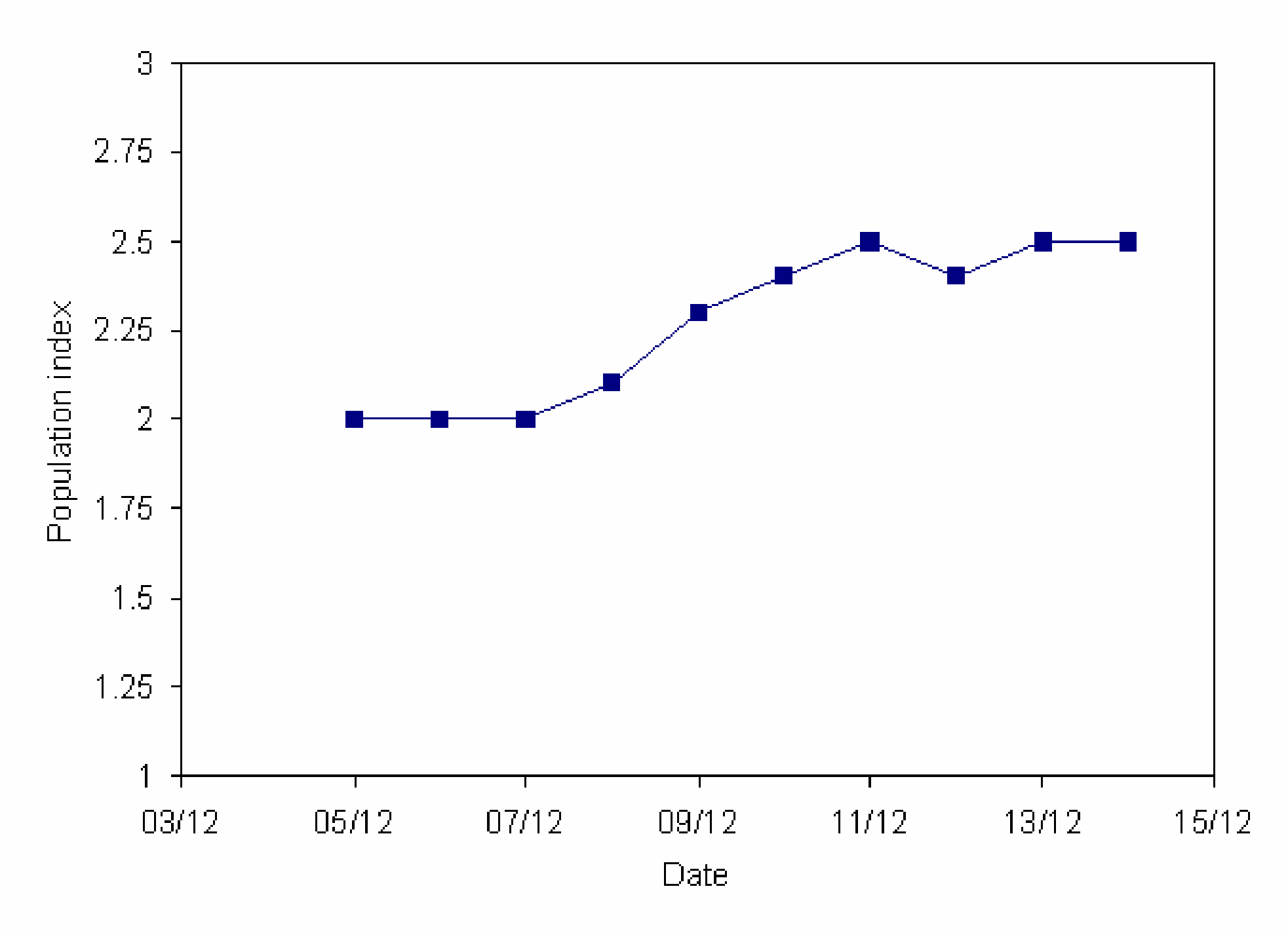

Figure 3. The population index of the Geminids in **2013** is shown as a function of date in December **2013**, based on measurements obtained by ourselves using video cameras in the Spanish meteor network. It illustrates that outside the main activity of the shower the population index was lower.

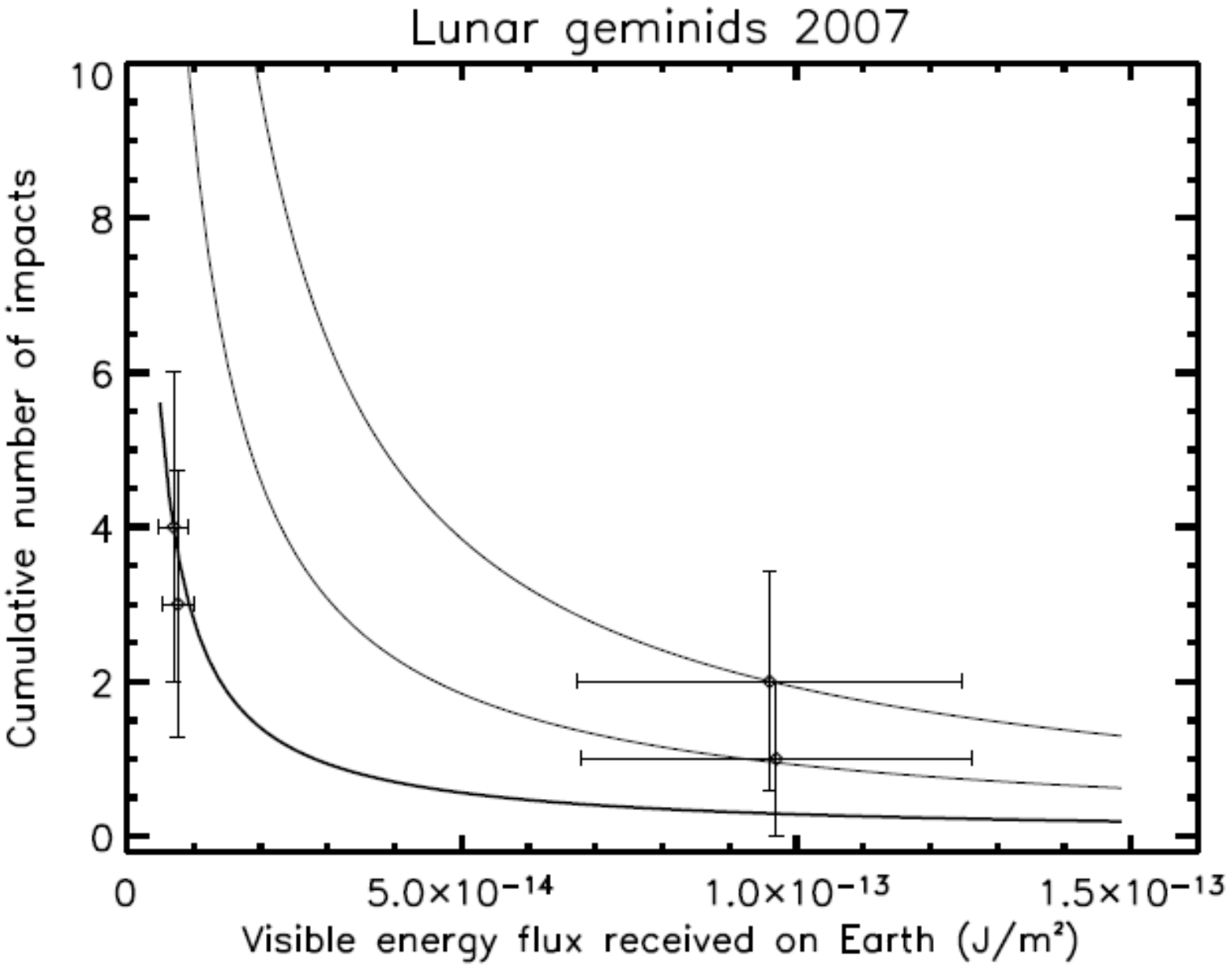

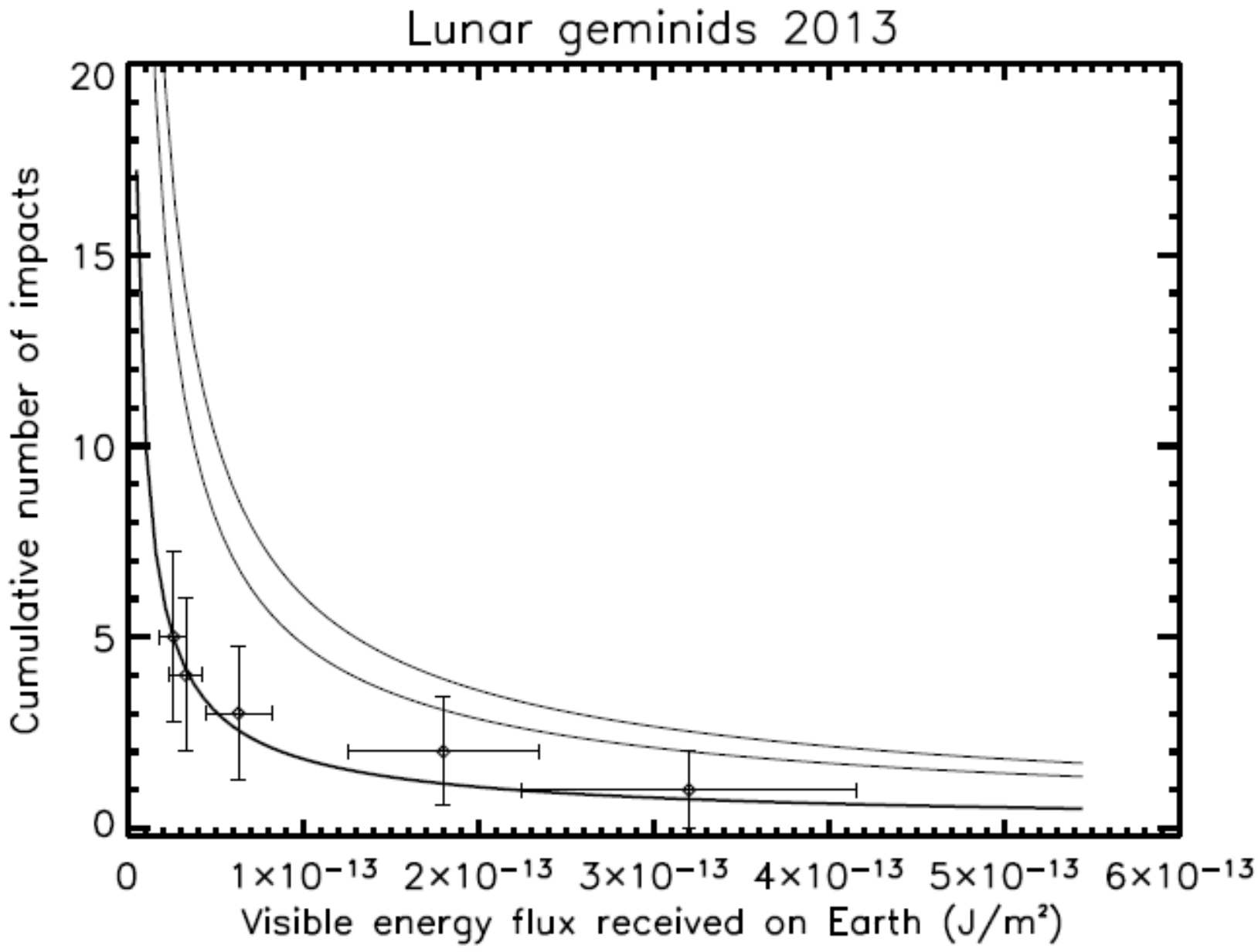

Figure 4. Cumulative number of Geminids impacts observed in 2007 (upper panel) and 2013 (lower panel) versus emitted energy. Diamond symbols represent the measurement whereas the lines represent equation (2) for different luminous efficiencies. In thick line we show the best fits, which correspond to $1.8 \cdot 10^{-3}$ and $2.4 \cdot 10^{-3}$ for 2007 and 2013 respectively. The two other lines in the upper plot correspond to $6 \cdot 10^{-3}$ and $1.2 \cdot 10^{-2}$ and the two other lines in the lower panel correspond to $8 \cdot 10^{-3}$ and $1.2 \cdot 10^{-2}$

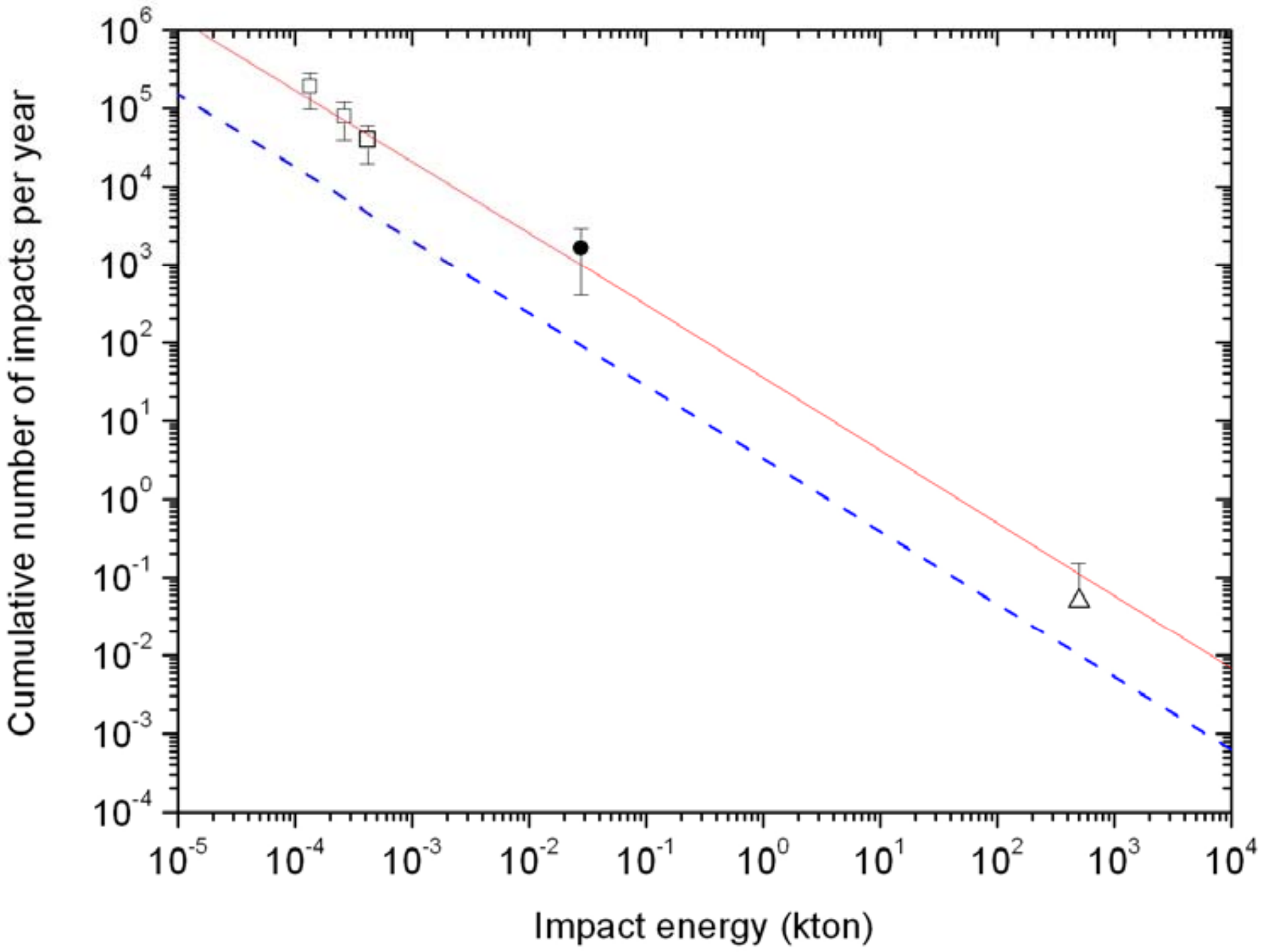

Figure 5. Cumulative number of impacts on Earth as a function of the kinetic energy of the impactors. This is a modified version of figure 6 in Madiedo et al. (2014a). The dashed blue line corresponds to the impact frequency reported in Brown et al. (2002), the squares correspond to the results derived from the lunar impact monitoring performed by Ortiz et al. (2006), modified for a luminous efficiency of $1.4 \cdot 10^{-3}$, the maximum that is compatible with lunar craters and other constraints. The filled black circle corresponds to the value derived in Madiedo et al. (2014a) also scaled to the $1.4 \cdot 10^{-3}$ luminous efficiency. The open triangle corresponds to the recent flux revision by Brown et al. (2013) for impactors delivering 500 kton. In red solid line we show the best fit to the three different groups of data.